\documentclass[twocolumn,showpacs,preprintnumbers,amsmath,amssymb]{revtex4}

\usepackage{graphicx}
\usepackage{dcolumn}
\usepackage{bm}

\begin{document}

\title{Spins, charges and currents at Domain Walls in a Quantum Hall Ising Ferromagnet.}

\author{L. Brey$^1$ and C. Tejedor$^2$}

\affiliation{$^1$Instituto de Ciencia de Materiales de Madrid (CSIC), 
Cantoblanco, 28049, Madrid, Spain.}

\affiliation{\centerline {$^2$Departamento~de~F\'{\i}sica~Te\'orica~de~la~Materia~ Condensada,~Universidad~Aut\'onoma~de~Madrid,~28049~Madrid,~Spain.}}

\begin{abstract}

We study spin textures in a quantum Hall Ising 
ferromagnet. Domain walls between ferro and unpolarized states at $\nu=2$ 
are analyzed with a functional theory supported by a microscopic calculation.
In a neutral wall, Hartree repulsion prevents the appearance of a fan phase
provoked by a negative stiffness. For a charged system, electrons become 
trapped as solitons at the domain wall. The size and energy of the solitons 
are determined by both Hartree and spin-orbit interactions. Finally, we 
discuss how electrical transport takes place through the domain wall.

\end{abstract}

\maketitle

Recently there is a great interest on the study 
of spin properties of quantum Hall states. For some filling factors, $\nu $, in 
the integer quantum Hall effect (QHE), a transition from a ferromagnetic (F)
to an unpolarized (U) ground state (GS) can be achieved by changing the ratio
between the cyclotron ($\hbar \omega _c$) and the Zeeman ($E_Z$) 
energies\cite{Giuliani}. Experimental evidence of this transition has 
been addressed recently\cite{Piazza,Poortere,Jungwirth0,Daneshvar}.
In the fractional QHE, transition between F and U states at  
filling factors $\nu=2/3$ and $2/5$ can be tuned by varying $E_z$ with 
respect the electron-electron interaction energy. Experimental indication 
of this transition has been 
reported\cite{Eisenstein,Engel,Kukushkin,Smet,Freytag,Eom,Kronmuler,Hashimoto} 
although the order of the transition is not 
clear\cite{Chakraborty,Apalkov,Murthy1}.

In this work we study a domain wall (DW) separating the U state from the 
F state, at $\nu$=2. At this filling factor, the F state has 
electrons occupying the Landau levels $n=0, 1$ with spin up. In the U
state, the electrons occupy the Landau level $n=0$ with the 
two spins orientations, i.e., the U state coincides with the singlet (S) state. 

At integer $\nu$, the phase transition between the uniform GS's can be described
by the  functional  $ {\cal F}=\alpha m_z + \beta m_z ^2$,
where ${\bf m} = ( {\bf m _{\perp}},m_z)$ 
is a unitary vector field, parallel to an isospin variable
which points to the positive (negative) $z$ direction when the 
GS is F (U)\cite{Jungwirth01}.
The phase transition occurs when $\alpha$=0. For $\nu$ being an integer 
greater than or equal to 2, $\beta $ is negative, indicating a first 
order phase transition 
between the $m_z$=1 and the $m_z$=-1 states: for this reason these systems 
are called quantum Hall Ising ferromagnets. The existence of hysteresis in
transport experiments is the smoke signal of the occurrence of 
a first order phase transition\cite{Piazza,Poortere,Jungwirth0,Daneshvar}.

In this work we present the following results:

i) We obtain a functional for describing isospin textures in  
the system at $\nu$=2. Due to the odd parity of the product of the  
Landau levels wave-functions participating in the F and S phases, we
find a negative stiffness for distortions of ${\bf m} _{\perp}$. This 
stiffness is not able to change the order of the phase transition at $\nu$=2. 

ii) By integrating out the transverse coordinate, we obtain a 
one-dimensional functional for describing spin textures at the DW. 
The adequacy of the functional is established by microscopic calculations.
This functional also has a negative stiffness, which  
could produce a fan spin texture in the transverse isospin component along 
the DW but, for GaAs quantum wells having widths of a few hundred $\AA$ 
and electron densities from $10^{11}$ to $10^{12} cm^{-2}$, 
Hartree repulsion prevents the formation of such topological structure.

iii) When the system is charged, extra electrons get trapped at the DW 
as topological excitations, solitons, with size 
and energy controlled by both the Hartree and spin-orbit (SO) interactions.  
The energy of the soliton controls the transport properties through the DW, 
the conductance being non-zero
only for finite SO coupling. This result solves the problem of the
spin conservation; transport through a DW implies a carrier spin flip 
something that can occur in presence of SO interaction.

{\em Energy functional for isospin textures.} 
The electron states of a two-dimensional electron gas confined in the $x-y$ 
plane and a magnetic field applied in the $z$-direction, are characterized by 
the Landau level index $n$, the degeneracy 
index $X$ and the spin $\sigma $. In the Landau gauge, $X$ is the momentum 
in the $y$ direction as well as the orbit center of the $x$-part of the 
wave-function. In this work, the magnetic length $\ell$ and the interaction 
$e^2/\epsilon \ell$ are the units of length and energy respectively.
In both the F and the S states, all the $|n=0, X, \sigma = \uparrow 
\rangle $ states are occupied and 
we consider them as electrically inert, being included in the vacuum.
The $\nu =2$ states are described by
\begin{equation}
\Psi \! = \! \prod _{X}
\! \left (  \! \cos { \theta (X)} c^\dagger _{X, \Uparrow} \! + \!  
\sin {\theta (X)}e ^ { i \psi ( X )  } c^\dagger _{X-G, \Downarrow} 
\! \right )|0 \rangle , 
\label{wavefunction}
\end{equation}
where $|0 \rangle$ is the vacuum, $c^\dagger$ are creation operators, 
$X$ runs over all possible states and the isospins $\Uparrow$ 
and $\Downarrow$ represent the states $n=1, \sigma = \uparrow$ and $n=0, 
\sigma = \downarrow$ respectively. In Eq.(\ref{wavefunction}) we only 
mix two isospins, since we suppose that $\hbar \omega _c$
is large enough for not producing Landau level mixing in the S and F 
phases\cite{Murthy2,Park}. Assuming that $\theta (X)$ and $\psi ( X )$ 
change slowly and $G$ is small, the unitary vector field corresponding 
to the state (\ref{wavefunction}) has the form, 
\begin{eqnarray}
m_z (x) & =  & \cos {2 \theta (x) } 
\nonumber \\
m_x (x,y)+i m_y (x,y) & = & \sin {2 \theta (x) }e ^ { i (\psi ( x )  +Gy)}. 
\label{field}
\end{eqnarray}
By computing the expectation value of the energy for the wave-function 
Eq.(\ref{wavefunction}),
we obtain the following energy functional for isospin textures:
\begin{eqnarray}
& & {\cal {F}}_{2D} \! = \! \alpha  \! \int \! d {\bf r} m_z( {\bf r} ) +
\beta \int \! d {\bf r} \, m_z ^2 ( {\bf r} )   +
{{\rho _{\parallel}} \over 2} \int d {\bf r}  \, 
\left (  \partial _{\mu} m _ z ({\bf r})  \right ) ^2
\nonumber \\ & & +
{{\rho _{\perp}} \over 2} \int \! d {\bf r} \, 
\left (  \partial _{\mu} \bf{ m }_ {\perp} ({\bf r})   \right ) ^2
+\hat {\rho} 
\int \!  d {\bf r} \,  
\left (  { \partial _{\mu}} ^2  \bf{ m }_ {\perp} ({\bf r})   \right ) ^2 
+V _{H}. 
\label{functional2D}
\end{eqnarray}
The coefficients are, 
\begin{eqnarray}
\alpha &  = &  \alpha _1 +
\left  ( E_Z - {{\hbar \omega _c} \over 2} \right ) {1 \over {2 \pi}} \, \, \nonumber \\
\alpha_1 & =  & {1 \over {8 \pi}} \left ( \Sigma_{0,0,0,0}
-  \Sigma_{1,1,1,1}
-  \Sigma_{1,0,1,0} \right )
\nonumber \\
\beta  & = &  {1 \over 16 \pi }  \left ( \Sigma _{0,0,0,0} +\Sigma _{1,1,1,1}
- 2 \Sigma _  {1,1,0,0} \right ) 
\nonumber \\
\rho _{\parallel}   &   = &  {1 \over 2} \left ( \rho  ^ {0,0}
+ \rho  ^ {1,1} \right )  \, \, \, \, \,   , \, \, \, \, \,
\rho _{\perp}   =    \rho  ^ {1,0} 
\label{alphas}
\end{eqnarray}
with
\begin{eqnarray}
& & \Sigma_{n,n_1,n_2,n_3}  =  - {1 \over S} \sum _{\bf q}
v({\bf q}) F_{n,n_1} ( {\bf q})
F_{n_2,n_3} (-{\bf q})
\nonumber \\
& & \rho  ^ {n,n_1}   =  
  {1 \over {2 \pi L}} \sum _{{\bf q}} { { q ^2} \over 4}
v({\bf q}) F_{n,n} ( {\bf q})
F_{n_1,n_1} ( -{\bf q})
\nonumber \\
& & {\hat {\rho}}    =  
{1 \over {48  \pi L}} \sum _{{\bf q }} { { q ^4} \over 4}
v({\bf q}) F_{n,n} ( {\bf q})
F_{n_1,n_1} ( -{\bf q})
\label{sigmas}
\end{eqnarray}
with $S$ and $L$ being the area and length (along the $y$-direction) of the 
system and $v({\bf q})$ the Fourier component of Coulomb interaction. 
For a strictly two dimensional system, the form factors are:
$F_{0,0}({\bf q})= e ^ {- q^2 / 4}$, 
$F_{1,1}({\bf q})=(1-q ^2/2)e ^ {- q^2 / 4}$ and
$F_{1,0}({\bf q})=(-q_y + i q_x) e ^ {- q^2 / 4}/\sqrt{2}$. 
The coefficients become:
$\alpha_1 = 3/ 32\sqrt{2 \pi}$, $\beta =-3/64\sqrt{2 \pi}$, 
$\rho _{\parallel}=11/128\sqrt{2 \pi} $, 
$\rho _{\perp}=-1/32\sqrt{2 \pi} $ and ${\hat {\rho}}=0.0035$. 

The novelty in this functional is the negative value of the transversal 
stiffness, $\rho _{\perp}$. In order to control the spatial variation
of ${\bf m} _{\perp}$, it is necessary to include in the expansion a higher 
derivative of ${\bf m} _{\perp}$. $\rho _{\perp} < 0$ due to the different 
parity of the $n=0$ and the $n=1$ Landau level wave-functions; in this way
$\rho ^ {1,0}$ and $\rho ^ {2,1}$ are negative whereas
$\rho ^ {2,0}$ and $\rho ^ {3,1}$ are positive. 
$\rho _{\perp}<0$ could produce intermediate helical phases between the 
F and the S states, however, at $\nu=2$ the magnitude of $\rho _{\perp}$ 
is not big enough for this occurrence.

Quantum Hall ferromagnets have the unique property that the topological 
charge is directly related to the electrical charge\cite{Sondhi,Fertig}. 
Therefore, we include in the functional Eq.(\ref{functional2D})
a Hartree term, $V_H$, representing the interaction between the charge 
densities $q({\bf r})= \varepsilon _{\mu,\nu} {\bf m} \cdot ( \partial 
_ {\mu} {\bf m} \times \partial _{\nu} {\bf m} )/8 \pi$ associated to 
the isospin texture. We use a standard\cite{Sondhi,Fertig} expresion
for $V_H$ including the semiconductor dielectric constant and
finite width of the quantum well. The F-S degeneracy occurs when 
$\alpha$=0, and the negative sign of $\beta$ indicates the first
order character of the transition. For $E_Z$=0 the phase transition occurs 
at $\hbar \omega _c=0.472$, which corresponds to an electron separation 
$r_s=2.12$. This justifies the use, in Eq.(\ref{wavefunction}), of just 
the $n=0$ and the $n=1$ Landau levels\cite{Murthy2}.

{\em Domain wall structure}
When $\alpha$=0, the S and the F states are degenerated, and 
disorder or finite temperature can produce DW's separating these GS's.
For studying  the structure of a DW, we assume $\alpha$=0 and
impose to the functional (\ref{functional2D}) the boundary conditions 
$m_z = \pm 1$ at $x = \pm \infty$. By doing that we obtain a DW thickness, 
$W_X$, of the order of $\ell$. 
In order to get a functional to describe a DW, we write 
${\bf m} _{\perp} = \sin {2 \theta (x)} 
(\cos{ \phi (y)}, \sin{ \phi (y)})$, and integrate 
in Eq.(\ref{functional2D}) over $x$ using a simple 
model in which $\theta (x)$ varies linearly through the DW, obtaining:
\begin{eqnarray}  
& & \Delta {\cal{F}}_{DW} (\phi (y))= \rho /2 \int dy (\partial_y\phi (y))^2 
\nonumber \\ & & + B\int dy [(\partial_y\phi (y))^4 +
(\partial_y^2\phi (y))^2] +V_{H}+\Delta {\cal{F}}_{SO}
\label{functionalDW}
\end{eqnarray}
where $\Delta {\cal{F}}_{SO}$ is a SO term 
that will be essential in the discussion below. Using the simple model 
$\theta (x)=\pi x/2W_X$ for $x<W_X$ and zero otherwise, the parameters in 
(\ref{functionalDW}) are $\rho = W_X\rho_{\perp}/2$ and $B=W_X{\hat {\rho}}/2$. 
However, the rapid change of $m_z(x)$ over a magnetic length,
raises some doubts on the validity of the functional 
(\ref{functional2D}) as a good starting point to obtain (\ref{functionalDW}).
Therefore, we have taken the alternative of performing a microscopic 
Hartree-Fock (HF) calculation\cite{Jungwirth} for describing DW's.
In Fig. 1 we plot the HF quasiparticle energies as
a function of the orbit guiding center. The chemical potential is located
at the gap energy. The reduction of the energy gap at the DW is an indication
of the loss of coherence of the wave-function. We find that 
$W_X$ is roughly $2 \ell$ and the energy per magnetic length 
of the DW is $0.0448$.
In the inset of Fig. 1 we plot the $z$-component of the unitary vector 
field, $m_z$, isospin as a function of the position.
At the center of the DW $m_z = 0$ and $m_{\perp}$ should be the unity. In
absence of spin-orbit coupling, the system has $U(1)$ symmetry and  the energy 
of the DW does not depend on global rotations of ${\bf m} _{\perp}$.
>From the HF results, we find that the functional (\ref{functionalDW}) 
is adequate for describing textures of ${\bf m} _{\perp}$ along the DW. The 
coefficients $\rho$ and $B$ for the terms with derivatives can be obtained 
from a fitting to the HF results. The ratio $\rho/B$ is the 
same than for the simple model above, but each coefficient has increased 
in a factor 3.7.
\begin{figure}
\includegraphics [clip,height=8.cm,width=9.cm]{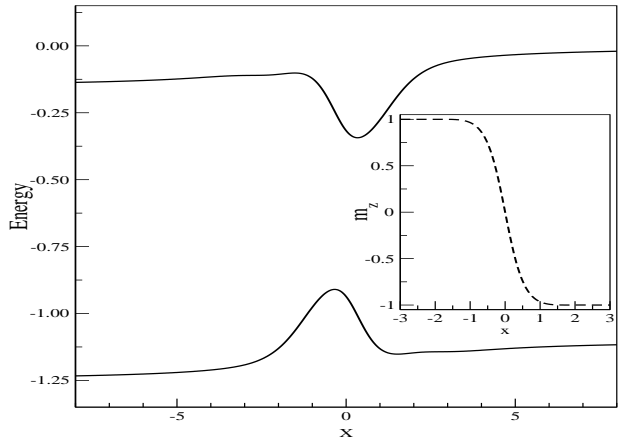}
\vspace{-1.5cm}
\caption{Dispersion relation (in units of $e^2/\epsilon \ell$) of the highest 
occupied and lowest unoccupied 
states in the region of the DW. The GS to the left is the F state while to 
the right is the S one. The inset shows the variation of the order 
parameter $m_z(x)$ at the DW. All the lengths are measured in units of $l$.} 
\end{figure}

Since $\rho<0$, the first term in (\ref{functionalDW}) tends to produce 
a rotation of the isospin along the DW. Although this rotation is limited by the 
second term, a fan phase could appear if one neglects any Hartree contribution 
as it is usually done\cite{Falko,Mitra}. However, a rotation implies the 
existence of electrical dipoles associated to oscillations of the 
topological charge (but with zero total topological charge) 
\begin{eqnarray}   
q({\bf r})=\frac{1}{4\pi}\partial_y\phi (y) \partial_x( m _z  (x)).
\end{eqnarray}  
The $V_H$ prevents the appearance of a fan phase induced
by the negative stiffness.
The Hartree repulsion of the charge density associated to the 
texture keeps the spin direction constant along the DW.
It must be stressed that the Coulomb term in Eq.(\ref{functionalDW}),
has the same dependence in derivatives of the field $\phi $ that the elastic 
term, and therefore it can not be neglected in the study of DW's.

We have also included in Eq.(\ref{functionalDW}) a SO term ${\cal{F}}_{SO}$.  
The SO interaction couples {\it directly} a state $\mid 0, X, \downarrow 
\rangle$ with a state $\mid 1, X, \uparrow \rangle$ \cite{Hanna} producing 
a Zeeman-like coupling to the isospin and an effective in plane magnetic 
field. Therefore, in our functional, SO is described by a term,
\begin{equation}
\Delta {\cal{F}}_{SO}=-\lambda _{SO} \int dy  \left ( cos \phi (y) -1 \right ) \, \, \, 
\end{equation} 
with $\lambda _{SO}=W_X \beta_{SO} /2^{3/2}\pi^2$ where $\beta_{SO}$ is the
bulk spin-orbit coupling \cite{Hanna}. 

{\em Charged domain wall}
The solutions of Eq.(\ref{functionalDW}) can be characterized by integers
which correspond to the total topological charge $Q_T$ of the solution.
$Q _T$ is the increase, in units of $2\pi$, of the phase $\phi (y)$ when going
from $- \infty$ to $+ \infty$. Hitherto, we have just considered solutions 
with $Q_T=0$. Let us now consider the solutions for $Q_T>0$.

Solutions of Eq.(\ref{functionalDW}) in the sector 
$Q_T=1$, are very important since the equivalence between topological and 
electrical charge allows the isospin textures to be the relevant charged 
excitations in the system\cite{Sondhi,Fertig}. In the presence of domains, 
charge excitations can be trapped in the walls forming confined isospin 
textures, which are solitons in the phase $\phi(y)$\cite{Falko}. 
Analytical expression for the soliton have been obtained, neglecting the 
Hartree interaction, in the case of positive stiffness\cite{Falko}.
In our case, the Hartree term is essential and we have not been able 
to obtain an analytic solution. For $Q_T=1$ we take a simplified shape for 
the soliton. In the sector of $Q_T=1$, we look for solitons
of size $\xi$ having a simple form $\phi(y)=2\pi y/\xi $,
within an interval of length $\xi$ and zero out of that interval.
The spin texture and the charge density of this soliton is shown schematically 
in Fig. 2. The size $\xi$ of the soliton is determined by the competition 
between the different terms of the functional; the Hartree and the 
quartic term ($B>0$) tend to make the texture large whereas the SO and the 
quadratic term ($\rho < 0$) try to make it small. 
Fig. 3 shows the energy and size of the charged wall as a function of 
$\lambda _{SO}$. For $\lambda _{SO}$=0, the  functional 
(\ref{functionalDW}) has $U(1)$ symmetry and the soliton has zero energy 
being extended to the whole wall, i. e. $\xi=L_y$. SO interaction ($\lambda 
_{SO} \neq 0$) reduces the soliton size. $\xi$ takes a value much smaller 
than $L_y$ and the energy of the excitation becomes finite. 
The energy of the soliton is the energy cost to add an electron to the DW. 
Since this value is smaller than the energy gap in the S and F phases, 
$\sim e^2 /\epsilon \ell$, we expect that extra charges 
in the system will become located at the DW.
The coupling $ \lambda _{SO}$ depends of the system characteristics as the 
DW width $W_X$. Typical values \cite{Hanna,Falko2} vary from   
$2 \times 10^{-4}$ to $8\times 10^{-4}$. In this range, the energy of the 
soliton is much smaller than the DW gap obtained in HF calculations 
(Fig. 1), $\sim  e^2 /\epsilon \ell$. The HF gap is dominated by 
exchange Coulomb interactions, and represents the excitation gap when the 
isospin order parameter is held fixed. The actual low-energy charge 
excitations come from fluctuations of the order parameter field. Once again, 
$V_H$ has been essential in the properties of the (in this case charged) DW.
\begin{figure}
\includegraphics [clip,height=8cm,width=10.cm]{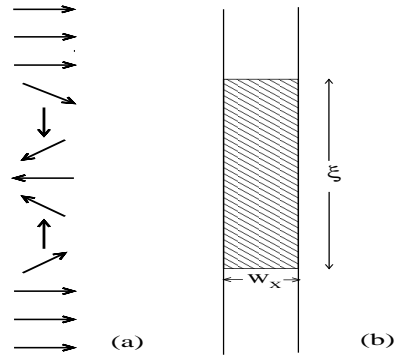}
\vspace{-1.5cm}
\caption{(a) Schematic behavior of the in-plane component ${\bf m}_{\perp}$ 
of the spin at the charged DW. (b) Schematic charge density (dashed region) 
of the soliton at the charged DW.} 
\end{figure}
\begin{figure}
\includegraphics [clip,height=8cm,width=9cm]{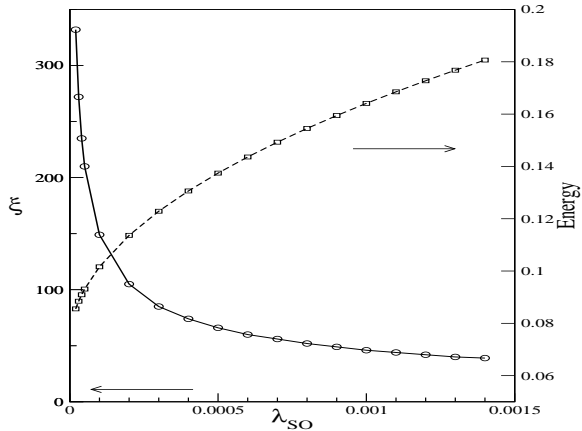}
\vspace{-1.5cm}
\caption{Size $\xi $ (left axis, in units of $l$) and 
energy (right axis, in units of $e^2/\epsilon \ell$) of the soliton in a 
charged DW as a function of the SO coupling.} 
\end{figure}

{\em Transport properties of the DW.}
Let us analyze the transport through a DW. If the chemical potential of the 
system, fixed by impurities or edge states, is located at the energy gap of 
the DW, no current can flow parallel to the DW\cite{Mitra2}. The only possibility for 
the carriers is to pass across the DW. On the contrary, when the
chemical potential resides in a band, there is a perfect unity transmission 
along the direction parallel to the wall and no carriers are passing through 
the DW. The same argument is valid for the charged excitations gap 
instead of that of the uncharged DW. The current through a DW separating a F 
from a U phase is different from zero if and only if the chemical potential 
lies on the charged excitation gap of the DW.

In the absence of SO coupling, there is not a gap for the charged excitations 
and, consequently, transport across the DW is not possible. This is in agreement 
with spin conservation arguments; when $\lambda _{SO}=0$, the spin is a good 
quantum number and no transport of charge through the DW is possible unless 
some other scattering mechanism is able to flip an electron spin.
The hyperfine coupling to nuclear spins has been sometimes invoked\cite{Smet}, 
but a non zero SO coupling is much more efficient to flip electron spins. 
Due to SO, the solution of the functional (\ref{functionalDW}) 
changes smoothly its isospin when going from one side to the other of the wall.
One electron with a given (real) spin can pass 
across the wall smoothly flipping its spin. The finite, due to SO 
coupling, energy of the soliton is rather small which means that very few 
electrons pass across the wall flipping their spins because a small gap reflects 
coupling between very few states at the two sides of the barrier\cite{Brey}. 
In other words, there is a small current passing across a domain wall with a 
large resistance. This explains the large resistance observed in different 
systems where domains exist\cite{Piazza,Poortere,Smet,Eom,Kronmuler,Hashimoto}. 

A final question to comment on is the role played by nuclear spins. 
Apart from the possible role played in the process of domain formation, 
nuclear spins are not needed, in our picture, in the process of carrier 
transport. However, due to the 
hyperfine interaction, nuclear spins will suffer a dynamic nuclear spin 
polarization within the electronic domains. This is very 
important because, if current is turned off for a while, as done in some 
experiments\cite{Hashimoto,Smet2}, the electrons in different domains 
immediately lose memory of theirs spins. However, nuclear spins relax so 
slowly in time that they serve as memory reservoirs of spin states and, if 
electronic current is reestablished after a while, the domains will reappear 
in exactly the same position they had before.

In summary, we study DW in a quantum Hall Ising ferromagnet at $\nu=2$ by 
means of a functional theory supported by a HF calculation.
In a neutral DW, Hartree repulsion prevents the appearance of a fan phase
provoked by a negative stiffness. When the system is charged, electrons are 
trapped as solitons at the DW. Hartree and SO interactions determine the 
energy and size of these solitons. Finally, a discussion of transport 
through the DW is presented.

We are indebted to L. Martin-Moreno for helpful discussions.
Work supported in part by MEC of Spain under contract No. PB96-0085.


\end{document}